\documentstyle[12pt]{article}
\input psfig.sty
\def\nn{\noindent}

\def\Re{{\cal R \mskip-4mu \lower.1ex \hbox{\it e}\,}}
\def\Im{{\cal I \mskip-5mu \lower.1ex \hbox{\it m}\,}}
\def\ie{{\it i.e.}}
\def\eg{{\it e.g.}}

\def\etal{{\it et al.}}
\def\ibid{{\it ibid}.}
\def\sub#1{_{\lower.25ex\hbox{$\scriptstyle#1$}}}

\def\to{\rightarrow}

\def\subw{_{\rm w}}
\def\mh{\ifmmode m\sbl H \else $m\sbl H$\fi}
\def\mch{\ifmmode m_{H^\pm} \else $m_{H^\pm}$\fi}
\def\mt{\ifmmode m_t\else $m_t$\fi}
\def\mc{\ifmmode m_c\else $m_c$\fi}
\def\mz{\ifmmode M_Z\else $M_Z$\fi}
\def\mw{\ifmmode M_W\else $M_W$\fi}
\def\mws{\ifmmode M_W^2 \else $M_W^2$\fi}
\def\mhs{\ifmmode m_H^2 \else $m_H^2$\fi}   
\def\mzs{\ifmmode M_Z^2 \else $M_Z^2$\fi}
\def\mts{\ifmmode m_t^2 \else $m_t^2$\fi}
\def\mcs{\ifmmode m_c^2 \else $m_c^2$\fi}
\def\mchs{\ifmmode m_{H^\pm}^2 \else $m_{H^\pm}^2$\fi}
\def\ztwo{\ifmmode Z_2\else $Z_2$\fi}
\def\zone{\ifmmode Z_1\else $Z_1$\fi}
\def\mtwo{\ifmmode M_2\else $M_2$\fi}
\def\mone{\ifmmode M_1\else $M_1$\fi}
\def\tb{\ifmmode \tan\beta \else $\tan\beta$\fi}
\def\xw{\ifmmode x\subw\else $x\subw$\fi}
\def\ch{\ifmmode H^\pm \else $H^\pm$\fi}
\def\lum{\ifmmode {\cal L}\else ${\cal L}$\fi}
\def\inpb{\ifmmode {\rm pb}^{-1}\else ${\rm pb}^{-1}$\fi}
\def\infb{\ifmmode {\rm fb}^{-1}\else ${\rm fb}^{-1}$\fi}
\def\epem{\ifmmode e^+e^-\else $e^+e^-$\fi}
\def\ppb{\ifmmode \bar pp\else $\bar pp$\fi}
\def\bsg{\ifmmode B\to X_s\gamma\else $B\to X_s\gamma$\fi}
\def\bsll{\ifmmode B\to X_s\ell^+\ell^-\else $B\to X_s\ell^+\ell^-$\fi}
\def\bstt{\ifmmode B\to X_s\tau^+\tau^-\else $B\to X_s\tau^+\tau^-$\fi}

\newskip\zatskip \zatskip=0pt plus0pt minus0pt
\def\matth{\mathsurround=0pt}

\def\atversim#1#2{\lower0.7ex\vbox{\baselineskip\zatskip\lineskip\zatskip
  \lineskiplimit 0pt\ialign{$\matth#1\hfil##\hfil$\crcr#2\crcr\sim\crcr}}}

\def\be{\begin{equation}}
\def\ee{\end{equation}}
\def\bea{\begin{eqnarray}}
\def\eea{\end{eqnarray}}
\renewcommand{\thefootnote}{\fnsymbol{footnote}}

\begin{document} \begin{titlepage} 
\rightline{\vbox{\halign{&#\hfil\cr
&SLAC-PUB-8001\cr
&November 1998\cr}}}
\vspace{1in} 
\begin{center}

{\Large\bf
Indirect Collider Signals for Extra Dimensions}
\footnote{Work supported by the Department of 
Energy, Contract DE-AC03-76SF00515}
\medskip

\normalsize 
{\large JoAnne L. Hewett} \\
\vskip .3cm
Stanford Linear Accelerator Center \\
Stanford CA 94309, USA\\
\vskip .3cm

\end{center}

\begin{abstract} 

A recent suggestion that quantum gravity may become strong near the weak
scale has several testable consequences.  In addition to probing for the new
large (submillimeter) extra dimensions associated with these theories via
gravitational experiments, one could search for the Kaluza Klein 
towers of massive gravitons which are predicted in these models and
which can interact with the fields of the Standard Model.
Here we examine the indirect effects 
of these massive gravitons being exchanged in fermion pair production 
in \epem\ annihilation and Drell-Yan production at hadron colliders.  
In the latter case, we examine a novel feature of this theory, which is the 
contribution of gluon gluon initiated processes to lepton pair production.  
We find that these processes provide strong bounds, up to several TeV, on the 
string scale which are essentially independent of the number of extra 
dimensions.  In addition, we analyze
the angular distributions for fermion pair production with spin-2 graviton
exchanges and demonstrate that they provide a smoking gun signal for 
low-scale quantum gravity which
cannot be mimicked by other new physics scenarios.  

\end{abstract}

\renewcommand{\thefootnote}{\arabic{footnote}} \end{titlepage} 


It has recently been suggested\cite{nima} that the hierarchy problem, \ie, 
the smallness of the ratio of the weak scale to the Planck scale ($M_{Pl}$),
may be avoided by simply removing the hierarchy.  In this case, gravitational
interactions become strong near the weak scale and take place mainly in $n$
new large spatial dimensions, known as the bulk.  Due to experimental
constraints, \eg, the width of the $Z$-boson, Standard Model (SM) fields
cannot propagate into the bulk and are forced to lie on a wall, or
3-dimensional brane, in the higher-dimensional space.  Gravity thus only 
appears to be weak in ordinary 4-dimensional space-time as we only 
observe its projection onto the wall.  The relation between the scales
where gravity becomes strong in the $4+n$ and 4-dimensional theories can be
derived from Gauss' Law and is given by
\be
M^2_{Pl}\sim r^n M^{2+n}_{eff}\,,
\ee
where $r$ is the size of the additional dimensions and $M_{eff}$ is the
effective Planck scale in the bulk.  The hierarchy dilemma is thus
resolved by taking $M_{eff}$ to be near a TeV, which yields $r\sim
10^{30/n-19}$ meters.  In this scenario, $n=1$ theories are automatically 
excluded as $r$ would be too large, while the case of $n=2$ with $r$ at a
sub-millimeter will be probed by future gravitational experiments\cite{grexp}.
In addition, it has been recently shown\cite{tye} that this framework can be 
embedded into string models, where the effective Planck scale can be
identified with the string scale $M_s$.
While we concentrate on this particular scenario, we note that there have 
been other interesting suggestions\cite{strings} for a low effective
Planck, or string, scale and for larger extra dimensions arising from string 
theory and Kaluza Klein models.

While this is a fascinating concept, what makes this theory really interesting 
is that it has testable consequences.  One manifestation of these theories
is the existence of a Kaluza Klein (KK) tower of massive gravitons which can
interact with the SM fields on the wall.  Here we examine the indirect effects 
of these massive gravitons being exchanged in fermion pair production 
in \epem\ annihilation and 
Drell-Yan production at hadron colliders.  As we will see below, these
processes provide strong bounds on the effective Planck scale which are
essentially independent of the number of extra dimensions.  In addition, we
quantify the extent to which the spin-2 nature of the graviton exchange is 
distinguishable from other potential new physics contributions to 
$\epem\to f\bar f$.  In the case of Drell-Yan production, we examine a
novel feature of this theory, which is the contribution of
gluon-gluon initiated processes to lepton pair production.  The direct
production of KK
excitations of the gauge bosons and Higgs fields at colliders has been
studied in \cite{kkprod}.

The effective theory below $M_{eff}$ consists of the SM fields and 
y-states\cite{nima} on the wall (the y-states are infinitely massive 
if the wall is rigid, or they can be Nambu-Goldstone
bosons if the translational invariance of the wall in the extra dimensions is
broken spontaneously), and gravity which
propagates in the full $4+n$ bulk.  The interactions of the y-modes are 
dependent on the specific dynamics of the brane\cite{rs} and we will not
consider them here.  The bulk metric can be written as
\be
G_{\hat\mu\hat\nu}=\eta_{\hat\mu\hat\nu}+{h_{\hat\mu\hat\nu}(x^\mu,x^a)\over
M_{eff}^{n/2+1}}\,,
\ee
where the indices $\hat\mu$ extend over the full $4+n$ dimensions, $\mu$ over
the $3+1$ dimensions on the wall, and $a$ over the $n$ bulk dimensions.  The
graviton field-strength tensor, $h_{\hat\mu\hat\nu}$, can be decomposed into
spin-2, 1, and 0 fields.  The interactions of these fields are given by
\be
\label{ndints}
\int d^{4+n}x\, T^{\hat\mu\hat\nu}{h_{\hat\mu\hat\nu}(x^\mu,x^a)
\over M_{eff}^{n/2+1}}\,,
\ee
where $T^{\hat\mu\hat\nu}$ is the symmetric, conserved stress-energy tensor in 
the bulk.  The induced metric on the wall 
is given by $G_{\mu\nu}(x^\mu,x^a=0)$ and
the interactions with the SM matter fields are obtained by decomposing
(\ref{ndints}) into the 4-dimensional states.  The bulk fields 
$h_{\hat\mu\hat\nu}$ appear as Kaluza-Klein towers in the 4-dimensional space 
arising from a Fourier analysis over the cyclic boundary conditions of the
compactified dimensions.  Performing this decomposition, we immediately
see that $T_{\mu a}=0$ and hence the spin-1 KK states don't
interact with the wall fields.  The scalar, or dilaton, states couple 
proportionally to the trace of the stress-energy tensor.  For interactions
with fermions, this trace is linear in the fermion mass, while for gauge
bosons it is quadratic in the boson mass.  Hence, the dilaton does not
contribute to the processes under consideration here.  

We thus only have to consider the interactions of the KK 
spin-2 gravitons with the SM fields.  All the gravitons in the KK tower,
including the massless state, couple in an identical manner.  Hence we
may use the couplings to matter as obtained in the case of linearized general
relativity\cite{gr}.  In this linearized theory, the matrix element for
$\epem\to f\bar f$ generalized for the case of $n$ massive graviton 
exchanges can be written as
\be
\label{mekk}
{\cal M}  =  {1\over M_{Pl}^2} \sum_n {T^e_{\mu\nu}P^{\mu\nu\lambda\sigma}
T^f_{\lambda\sigma}\over s-m^2_{gr}[n]} \,,
\ee
where the sum extends over the
KK modes.  $P_{\mu\nu\lambda\sigma}$ 
represents the polarization sum of the product of two graviton fields
and is given in \cite{gr}.  The terms in the polarization sum that are 
quadratic and quartic in the transferred momentum do not contribute
to the above matrix element since $T_{\mu\nu}$ is
conserved.  Likewise, the terms which go as $\eta_{\mu\nu}
\eta_{\lambda\sigma}$ lead to terms proportional to $T_\mu^{e\mu}
T_\lambda^{f\lambda}$ which vanish in the limit of zero electron mass.
The remaining terms are $P_{\mu\nu\lambda\sigma}={1\over 2}[\eta_{\mu\lambda}
\eta_{\nu\sigma}+\eta_{\mu\sigma}\eta_{\nu\lambda}-\eta_{\mu\nu}
\eta_{\lambda\sigma}]$ and are exactly
those present in the massless graviton case; they are thus universally 
applicable to all of the states in the KK tower.  Since the spacing of
the KK states is given by $\sim 1/r$, the sum over the states in (\ref{mekk}) 
above can be approximated by an integral which is log divergent for $n=2$
and power divergent for $n>2$.  A cut-off must then be applied to regulate
these ultraviolet divergences, and is generally taken to be the scale of the 
new physics.  For $n>2$ it can be shown\cite{nima,joe} that the dominant
contribution to this integral is
of order $\sim M^2_{Pl}/M_s^4$, where we have taken the cut-off to be the 
string scale, while for $n=2$ this result is multiplied by a factor of order 
$\log(M_s^2/E^2)$, where $E$ is the center-of-mass energy of the process
under consideration.  The exact computation of this integral can only be 
performed with some knowledge of the full underlying theory.  Combining these 
results yields the matrix element
\bea
\label{ffff}
{\cal M} & =  & {\lambda\over M_s^4}  \left\{  \bar e(p_1)\gamma_\mu e(p_2)
\bar f(p_3)\gamma^\mu f(p_4)(p_2-p_1)\cdot (p_4-p_3)\right.\\
& & \left. \bar e(p_1)\gamma_\mu e(p_2)\bar f(p_3)\gamma_\nu f(p_4)
(p_2-p_1)^\nu(p_4-p_3)^\mu\right\}\,.\nonumber
\eea
Here, the momentum flow is defined with $p_{1,2}$ into the vertex and
$p_{3,4}$ outgoing.  Note that graviton exchange is $C$ and $P$ conserving,
and is independent of the flavor of the final state.
The coefficient $\lambda$ is of ${\cal O}(1)$ and cannot 
be explicitly calculated without knowledge of the full quantum gravity theory.  
It is dependent on the number of extra dimensions, how they are compactified,
and is in principle a power series in $s/M_s^2$.  However, we neglect this
possible energy dependence in $\lambda$ and note that the limits obtained 
here, which go as $|\lambda|^{1/4}$, are only very weakly dependent on its
precise value and hence on the specific model realization.  In principle the
sign of $\lambda$ is undetermined and we examine the constraints that can be
placed on $M_s$ with either choice of signs.

The angular distribution for $\epem\to f\bar f$ with massive fermions is
then calculated to be
\bea
\label{dsdz}
{d\sigma\over dz} & = & N_c{\pi\alpha^2\over 2s}\beta\left\{ P_{ij}
\left[A^e_{ij}A^f_{ij}(1+\beta^2z^2)+2\beta B^e_{ij}B^f_{ij}z+
A^e_{ij}C^f_{ij}(1-\beta^2)\right]\right.
\nonumber\\
& & -{\lambda s^2\over 2\pi\alpha M_s^4}P_i\left[2\beta^3z^3v^e_iv^f_i
-\beta^2(1-3z^2)a^e_ia^f_i\right] \\
& & \left. +{\lambda^2s^4\over 16\pi^2\alpha^2M_s^8}\left[
1-3\beta^2z^2+4\beta^4z^4-(1-\beta^2)(1-4\beta^2z^2)\right]\right\} \,,
\nonumber
\eea
where the indices $i,j$ are summed over $\gamma$ and $Z$ exchange, $z=\cos
\theta$, $P_{ij}$ and $P_i$ are the usual propagator factors (defined in \eg,
\cite{leptos}), $\beta=(1-4m_f^2/s)^{1/2}$, $A^f_{ij}=(v_i^fv_j^f
+a_i^fa_j^f)\,, B^f_{ij}=(v_i^fa_j^f+v_j^fa_i^f)\,, C^f_{ij}=(v_i^fv_j^f
-a_i^fa_j^f)$, and $N_c$ represents the number of colors of the final state.
In the case of Bhabha scattering, $t$- and $u$-channel graviton
exchanges will also be present.
If polarized beams are available a z-dependent Left-Right 
asymmetry can also be formed:
\bea
\label{alr}
A_{LR}(z) & = & P_{ij}\left[ 
B^e_{ij}A^f_{ij}(1+\beta^2z^2)+2\beta A^e_{ij}B^f_{ij}
z+B^e_{ij}C^f_{ij}(1-\beta^2)\right]/D\\
& & -{\lambda s^2\over 2\pi\alpha M_s^4}
P_i\left[2\beta^3z^2a^e_iv^f_i-\beta^2(1-3z^2)v^e_ia^f_i\right]/D\,,\nonumber
\eea
where $D$ is given by the curly bracket in (\ref{dsdz}) above.
Note that the total cross section and integrated left-right asymmetry are
{\it unaltered} by graviton exchanges, independently of fermion flavor 
up to terms of order $s^4/M_s^8$, and hence
only the angular distributions for these quantities will be sensitive to these
new exchanges.  This is not the case for other new physics scenarios\cite{tgr}
which also have indirect contributions to $\epem\to f\bar f$ via new particle
exchange, such as those with additional neutral gauge bosons or with scalar
exchange in the $s/t$-channels as in \eg, 
supersymmetry with R-parity violation.  In
general, these other scenarios affect the total integrated quantities as
well as the angular distributions in a flavor dependent manner.  In addition,
the shape of the angular distributions for spin-2 exchange is unique and
provides a smoking gun signature for graviton exchange.

\begin{figure}
\centerline{
\psfig{figure=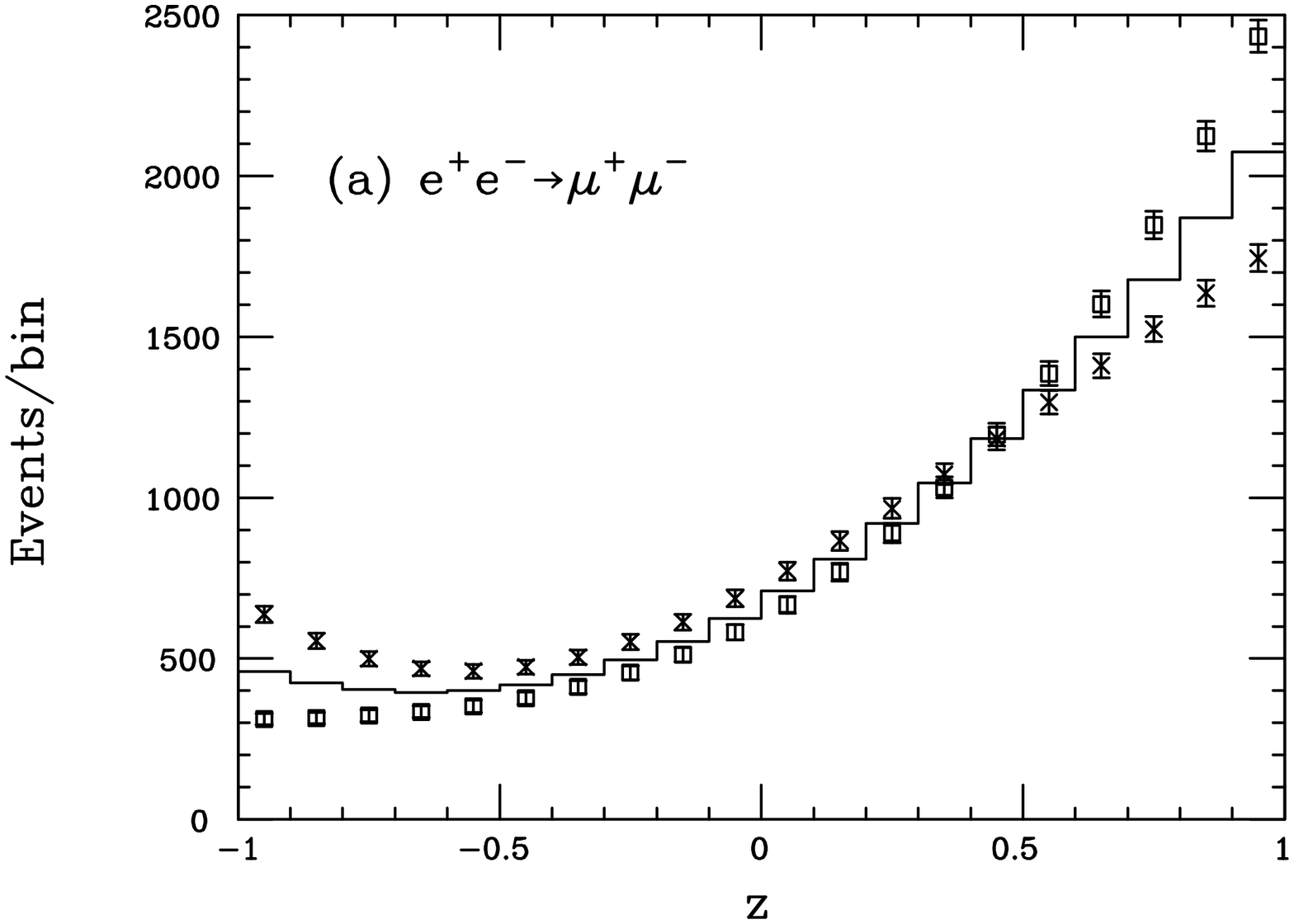,height=7.cm,width=8.5cm,angle=-90}
\hspace*{-5mm}
\psfig{figure=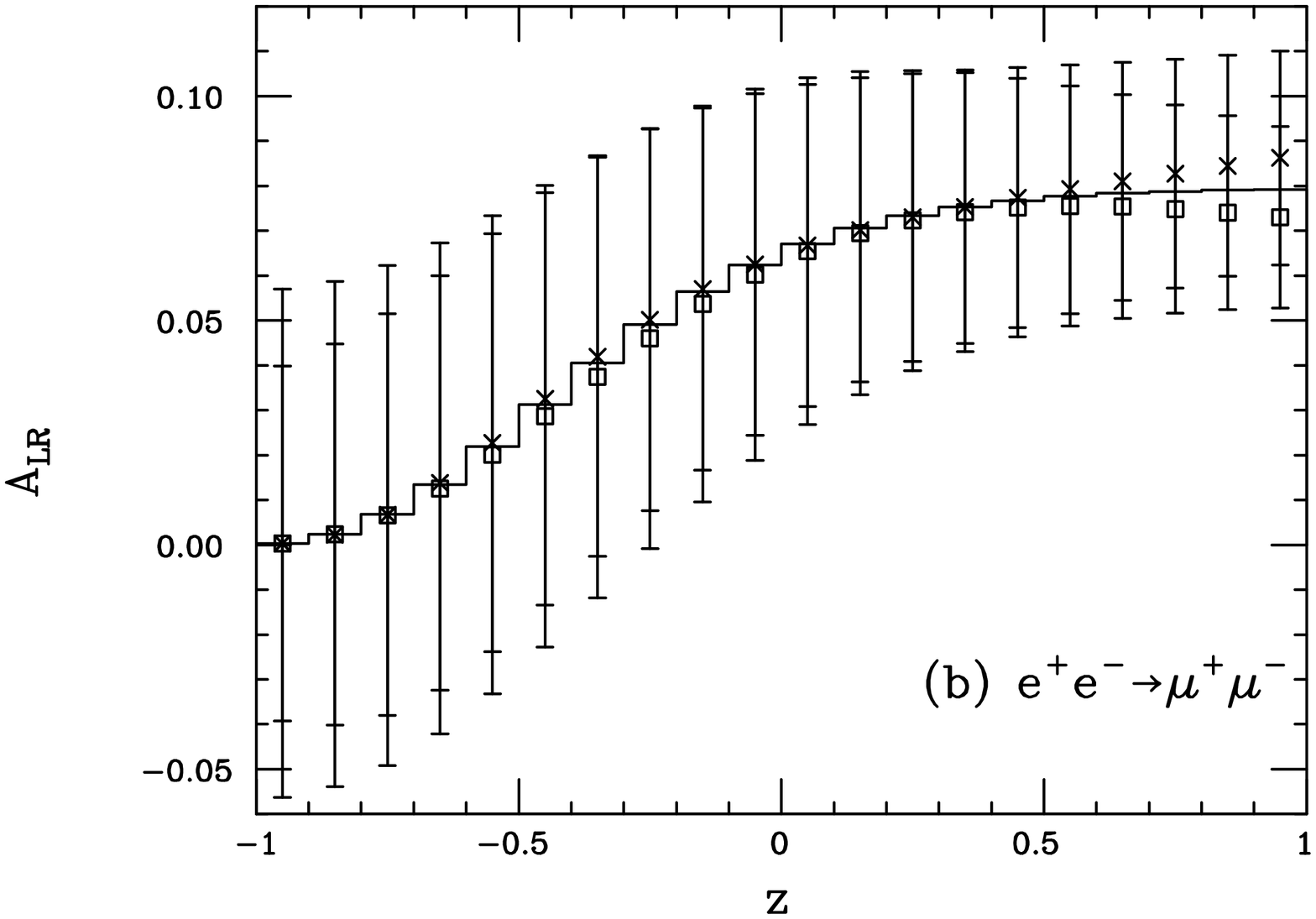,height=7.cm,width=8.5cm,angle=-90}}
\vspace*{-0.75cm}
\centerline{
\psfig{figure=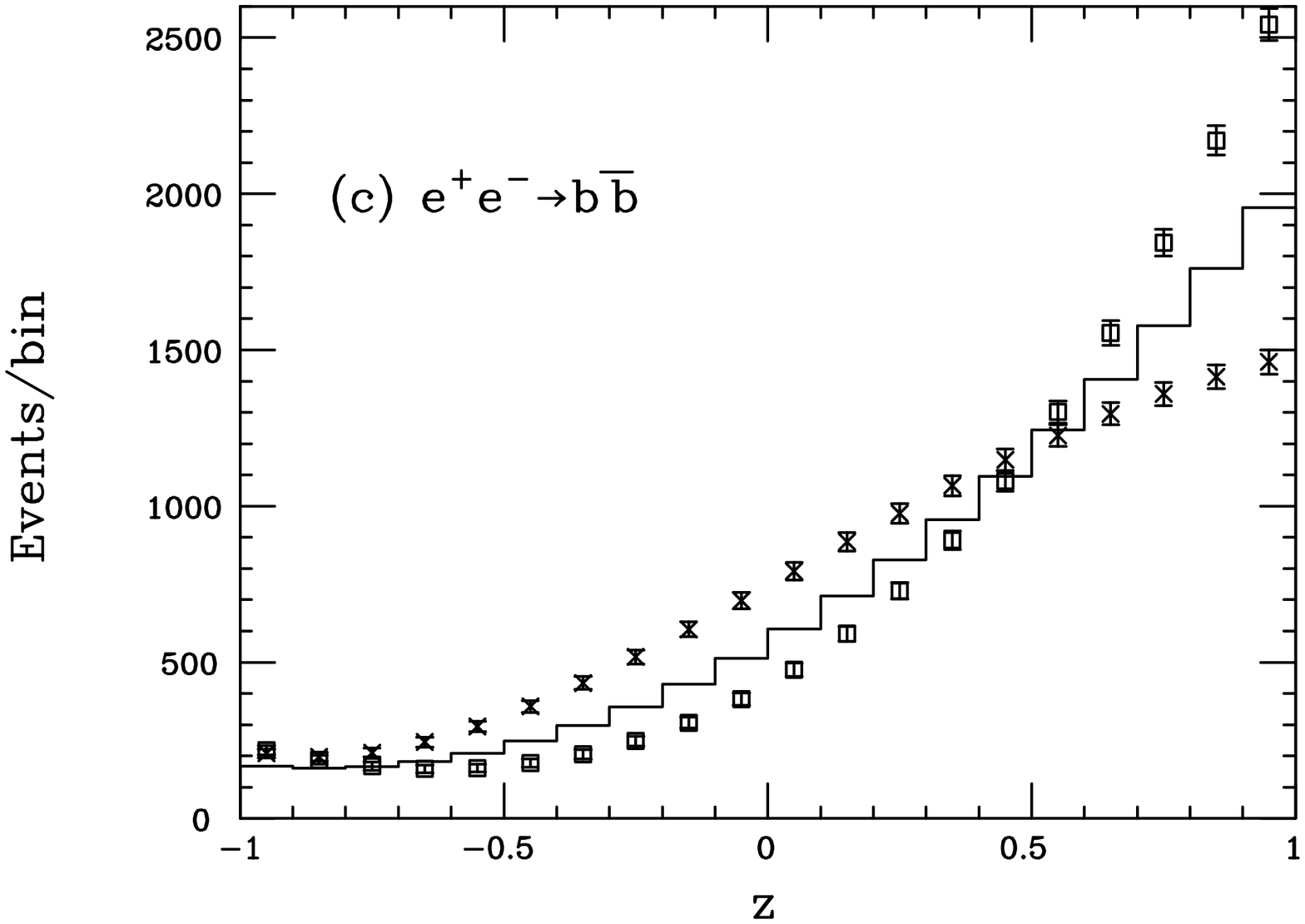,height=7.cm,width=8.5cm,angle=-90}
\hspace*{-5mm}
\psfig{figure=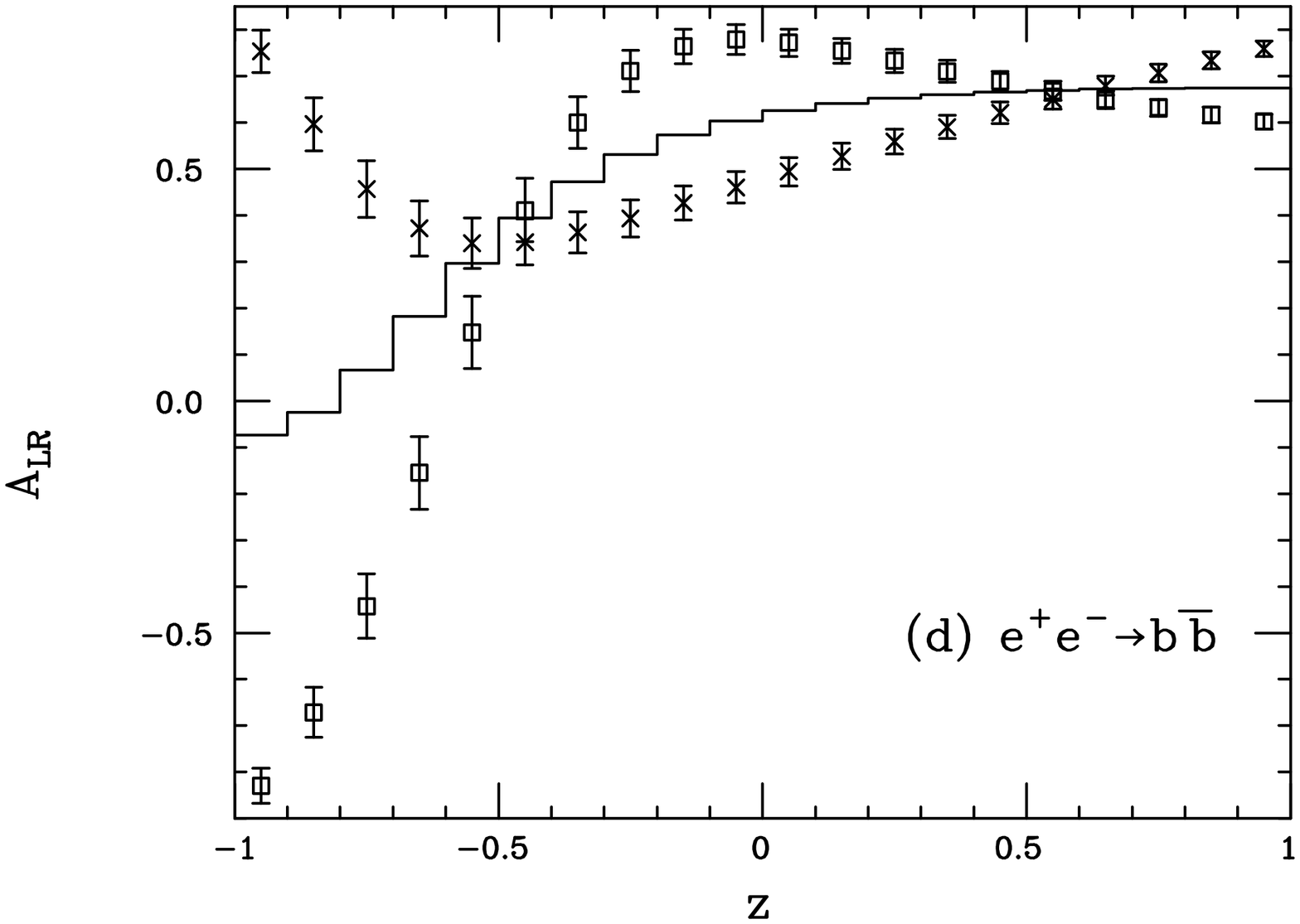,height=7.cm,width=8.5cm,angle=-90}}
\vspace*{-0.75cm}
\centerline{
\psfig{figure=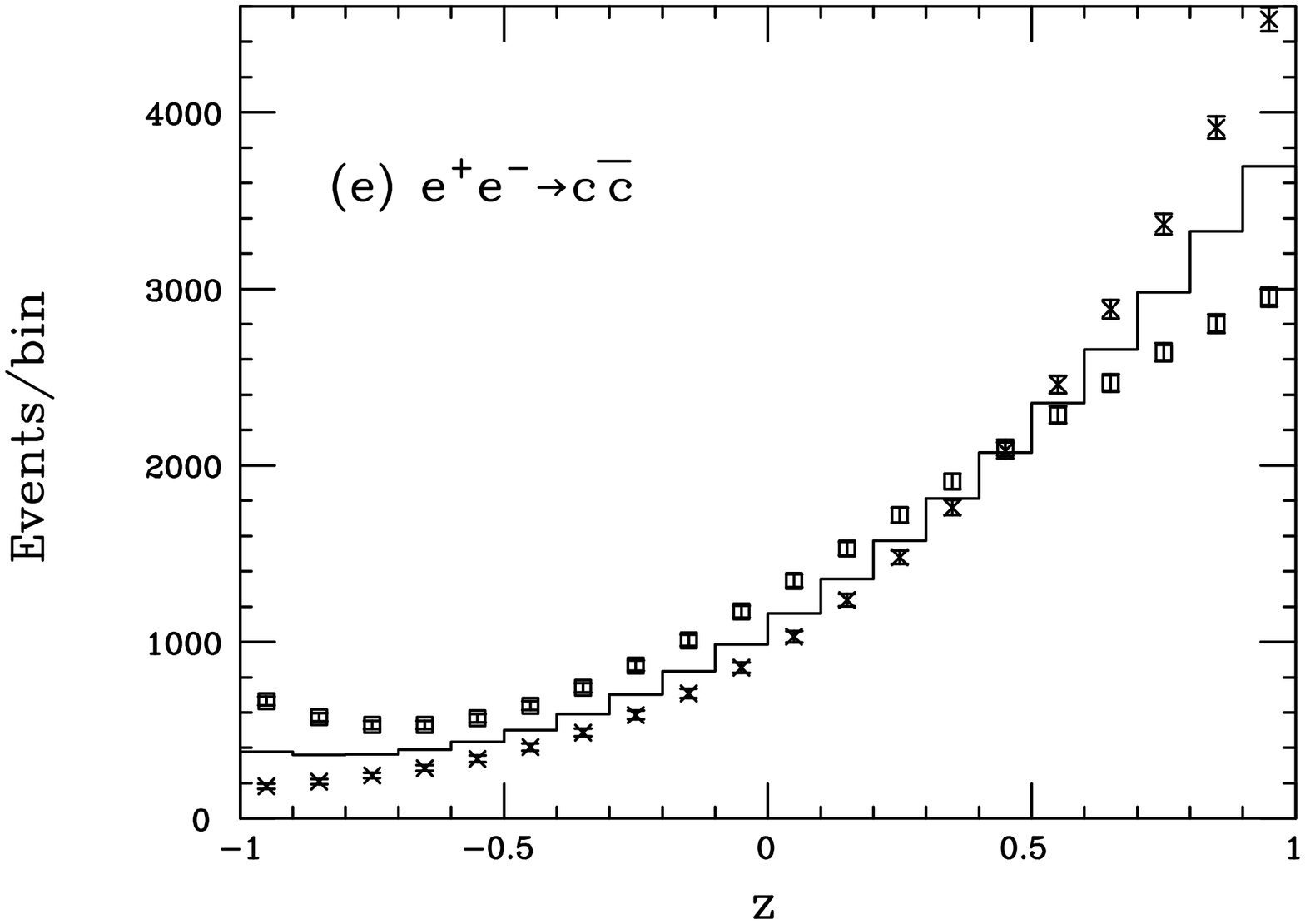,height=7.cm,width=8.5cm,angle=-90}
\hspace*{-5mm}
\psfig{figure=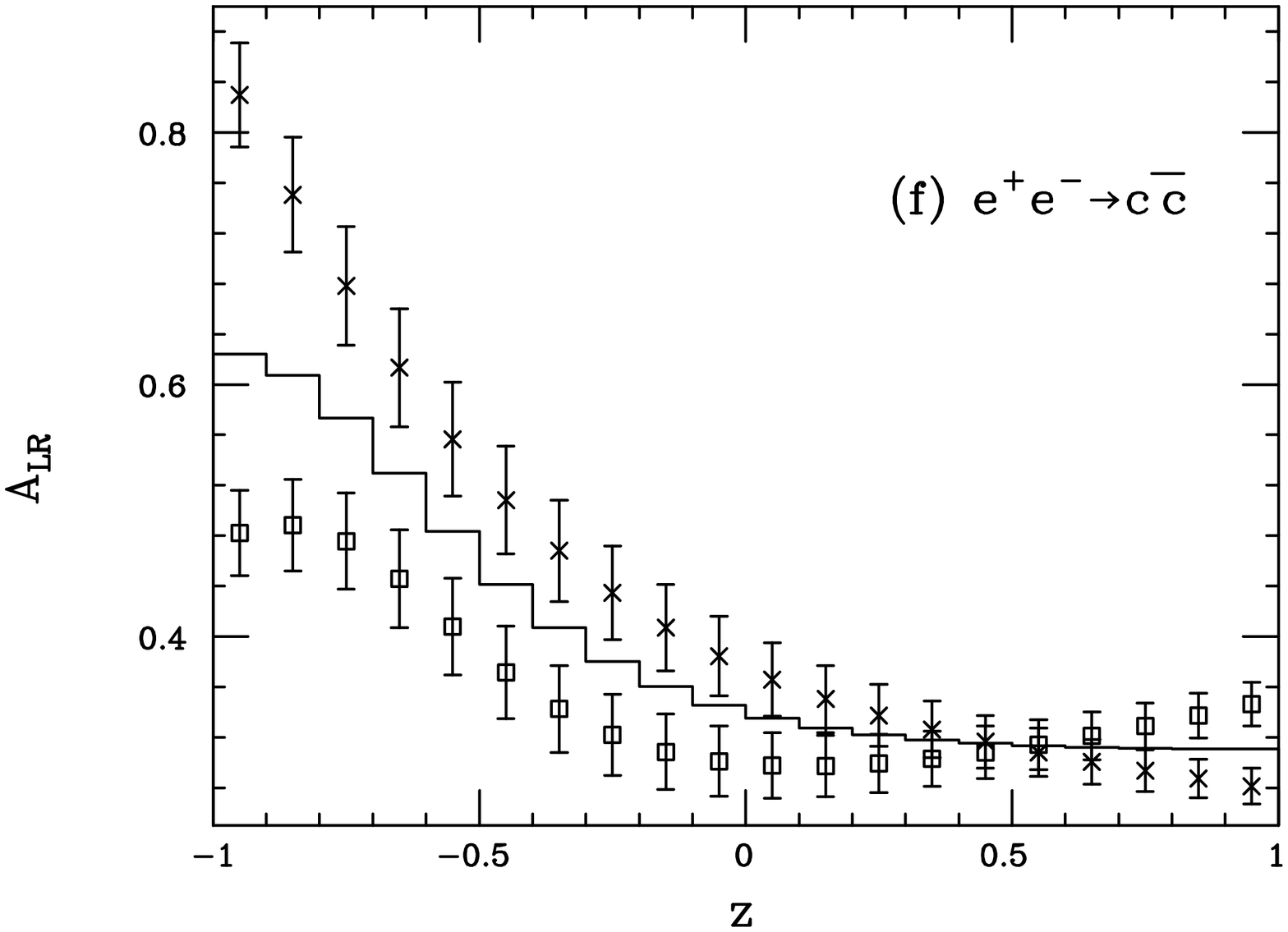,height=7.cm,width=8.5cm,angle=-90}}
\vspace*{-0.75cm}
\caption{Bin integrated angular distribution and $z$-dependent Left-Right 
asymmetry
for $\epem\to \mu^+\mu^-\,, b\bar b\,, c\bar c$.  In each case, the solid 
histogram represents the SM, while the `data' points are for $M_s=1.5$ TeV 
with $\lambda=\pm 1$.  The error bars correspond to the statistics in
each bin.}
\label{edists}
\end{figure}

\begin{figure}[t]
\centerline{
\psfig{figure=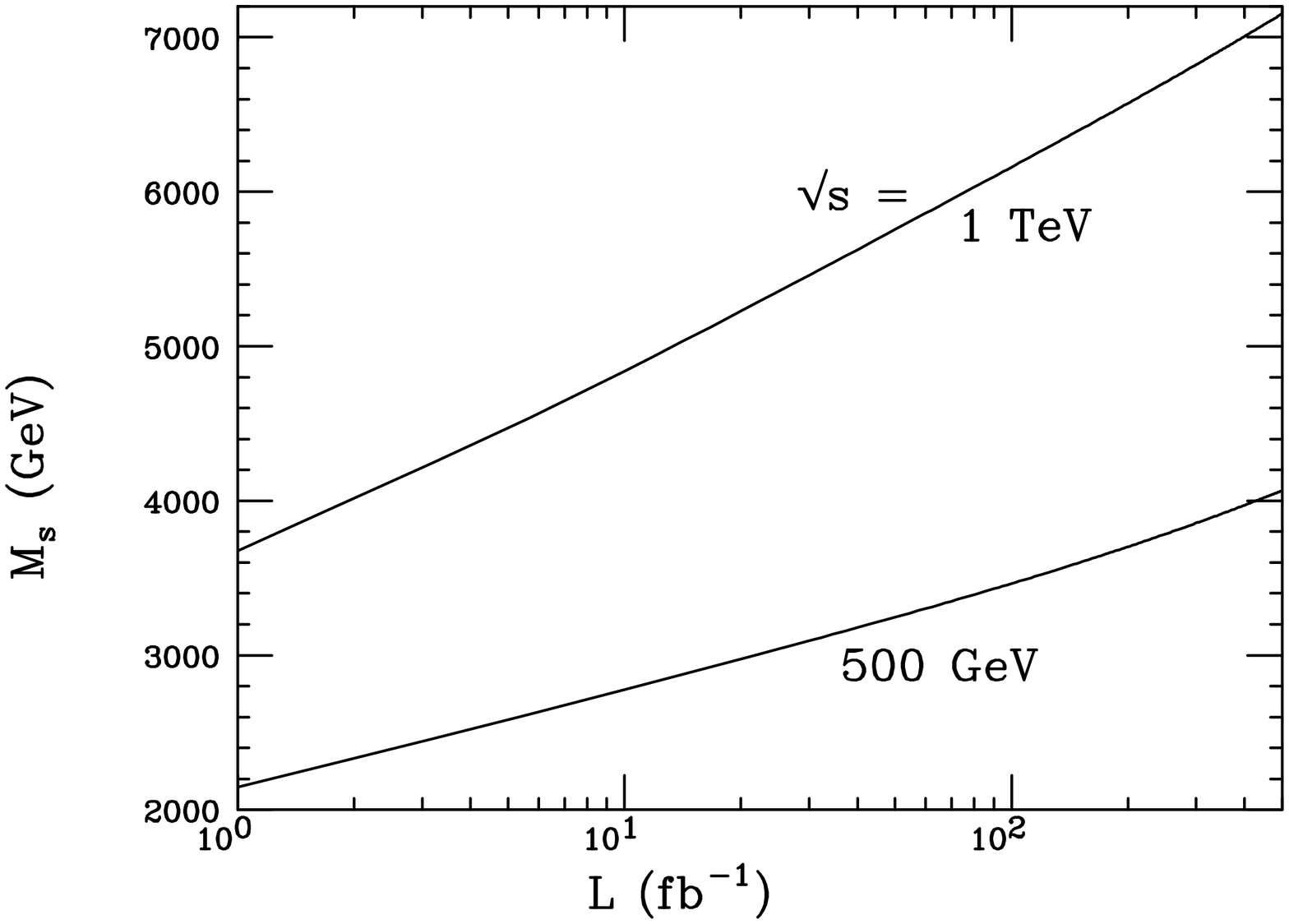,height=8.cm,width=8cm,angle=-90}
\hspace*{-5mm}
\psfig{figure=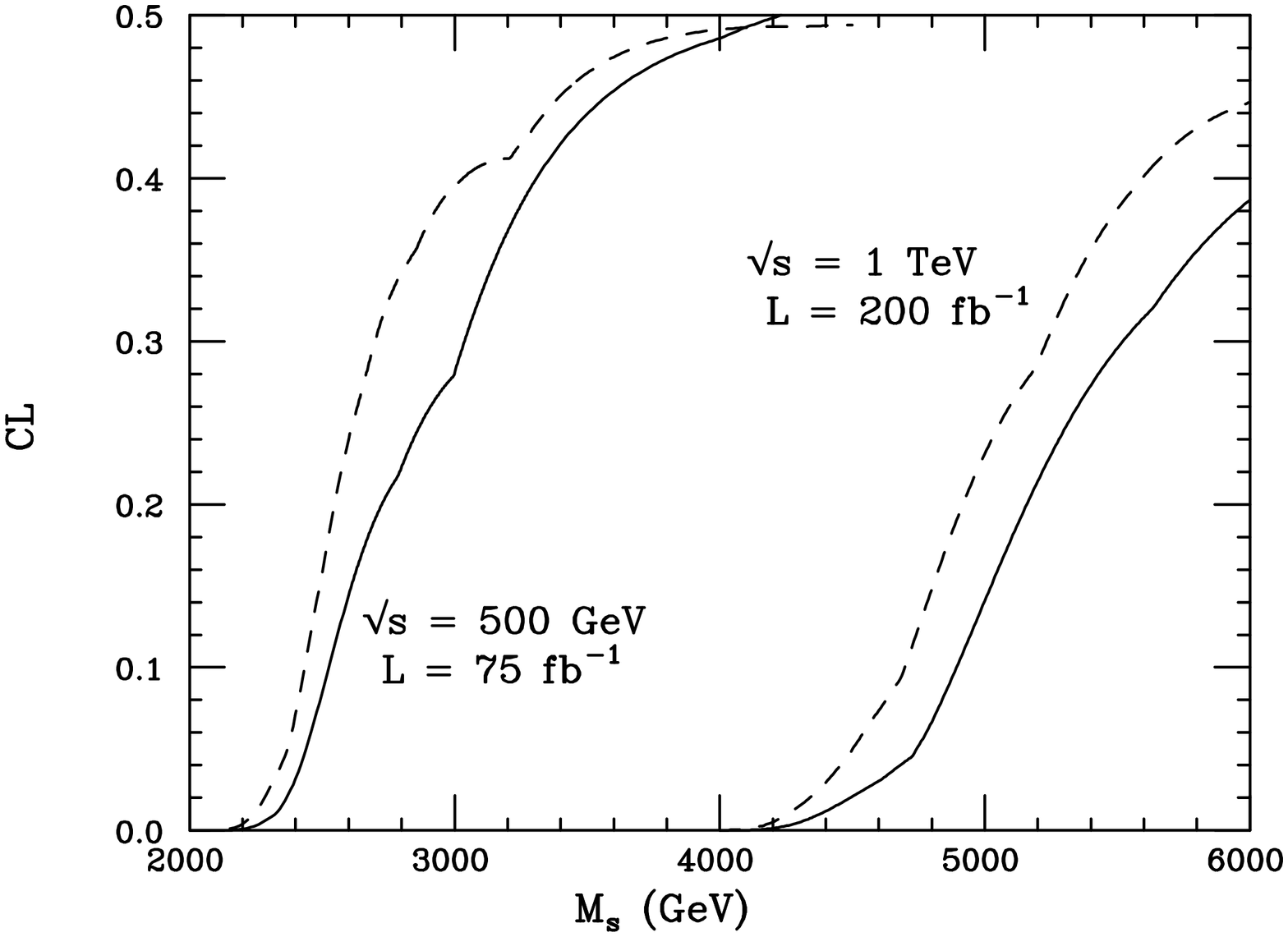,height=8.cm,width=8cm,angle=-90}}
\vspace*{-1cm}
\caption{Left: 95\% C.L. search reach for the string scale as a function of
integrated luminosity at \epem\ colliders with center-of-mass energy as
labeled.  Right: The percentage confidence level as a function of the string
scale for a fit to the `data' of
Fig. \ref{edists} assuming the angular distributions take the form expected
in the case of new gauge boson exchange.  The assumed center-of-mass energy and
luminosity is as labeled, and the dashed and solid curves in each case
correspond to the choice $\pm\lambda$.}
\label{eeres}
\end{figure}

The bin integrated angular distributions are displayed in Fig. \ref{edists} 
for $\mu^+\mu^-\,, b\bar b$, and $c\bar c$ final states with $\sqrt s=500$ GeV.
Here the solid histogram corresponds to the SM expectations and the `data' 
points represent the case with graviton exchanges with $M_s=1.5$ TeV.  The
two sets of data points (squares and x's) correspond to the two choices of
sign for $\lambda$.  The errors on the data points represent the statistics
in each bin for an integrated luminosity of $75\infb$.  Here, we have 
assumed a 60\% heavy quark tagging efficiency for $b$ and $c$ corresponding to 
what is expected\cite{djackson} to be achieved at high energy linear colliders,
we have taken the electron beam polarization to be  90\%, 
employed a $10^\circ$ angular cut around the beam pipe (to remove backgrounds
from the interaction region), and included the effects of initial state 
radiation.  We see that each of these distributions, with the exception of 
$A_{LR}(z)$ for $\mu$'s, provides a statistically significant signal for the
graviton exchanges.  Note that the deviations from the SM for the $b$-quark 
final state are particularly outstanding.  
From (\ref{dsdz}) we see that the expected shape of the angular distribution
for the SM, (or for any spin-1 exchange) goes as $\sim (1+z^2)$ and it is
clear by eye that the spectrum with the graviton exchanges do not have this
parabolic shape.  Unfortunately $A_{LR}(z)$ for
leptonic final states is numerically small and hence relatively poorly
determined and will carry little statistical weight in our analysis.
Summing over $e\,, \mu\,, \tau\,, b\,, c$ and $t$ final states (employing
a 60\% reconstruction efficiency for the case of top-quarks), including the
$\tau$ polarization asymmetry, and performing the usual $\chi^2$ 
analysis\cite{snow} results in the 95\% C.L. search reach shown
in Fig. \ref{eeres}(a) as a function of luminosity with center-of-mass 
energies as indicated.  Note that the effects of string scales up to 
$6\sqrt s$ (for canonical luminosity values at linear colliders) are
discernable.  Performing this same procedure for LEP II, but excluding the
$A_{LR}(z)$ observable since polarized beams are not available, excluding
top final states, and using heavy quark tagging efficiencies applicable for
the LEP II detectors, we find that string scales up to $M_s=985$ GeV are
excluded from present data (taken to be 200\inpb\ per detector at $\sqrt s=189$
GeV), and that this reach may be extended to $M_s=1.14$ TeV with 2.5\infb\
summed over all 4 detectors with $\sqrt s=195$ GeV.  
The 95\% C.L. search reaches for design luminosity at these machines
are constrasted in Table \ref{searchres}.
Note that these
constraints are actually placed on the quantity $|\lambda|^{-1/4}M_s$.  We
find that the difference in the search reach due to the sign ambiguity in
$\lambda$ is only a few GeV.

We note that the reaction $\epem\to gg$ can also be mediated by graviton
exchanges and will affect 2-jet production in \epem\ annihilation.  However,
since this reaction does not interfere with the 
four-fermion matrix element, it is suppressed by $1/M_s^8$ 
and hence will have little effect on the overall 
search reach in \epem\ collisions.

Next, we quantify the extent to which these spin-2 exchanges are distinguishable
from other new physics sources.  As an example, we perform a fit to the 
`data' shown in
Fig. \ref{edists} assuming that the unpolarized and polarized angular 
distributions take the forms $A(1+z^2)+Bz$ and $[C(1+z^2)+Dz]/[A(1+z^2)+Bz]$,
respectively, where $A\,, B\,, C$, and $D$ represent
arbitrary constants to be determined from the fit.  These forms are what
would be expected in the case of new vector boson exchange.
For the angular distribution, 
we include $e\,, \mu\,, \tau\,, b$ and $c$ final states (the top-quark is
excluded as its mass effects would alter the constants $A$ and $B$), while for
$A_{LR}(z)$ we only include $b\bar b$ and $c\bar c$ production
as the leptonic final states carry no statistical weight for this observable.
This corresponds to 88 degrees of freedom in the fit.  The value of $\chi^2$
per degree of freedom is computed and the confidence level of the fit is
presented in Fig. \ref{eeres}(b) as a function of the string scale.  We see
that the quality of the fit is quite poor for string scales up to $\sim 5
\sqrt s$, which is almost up to the discovery limit.
This demonstrates that spin-2 graviton exchanges are
easily separated from that of new vector bosons.  Similar studies can also
be performed for comparison with new scalar exchange\cite{tgr}.

\nn

\begin{table}
\centering
\begin{tabular}{|c|c|c|c|} \hline\hline
  & $\sqrt s$ (TeV) & ${\cal L}$ (\infb) & $\lambda=\pm 1$  \\ \hline
LEP II & 0.189 & 0.8 & 0.98 \\
       & 0.195 & 2.5 & 1.1  \\ \hline
Linear Collider     & 0.5   & 75  & 3.4 \\ 
       & 0.5   & 500 & 4.1  \\ 
       & 1.0   & 200 & 6.6  \\ \hline
Tevatron & 1.8 & 0.11 & 0.99 \\
         & 2.0 & 2    & 1.3 \\
         & 2.0 & 30   & 1.7 \\ \hline
LHC      & 14  & 10   & 5.2 \\
         & 14  & 100  & 6.0 \\ \hline\hline
\end{tabular}
\caption{95\% C.L. search reach for the string scale in TeV 
for various colliders with center-of-mass energies and integrated luminosities
as indicated.}
\label{searchres}
\end{table}

We now examine the case of lepton pair production in hadronic collisions.
The subprocess contribution of the graviton exchanges to ordinary 
Drell-Yan production is essentially
given by Eq. (\ref{dsdz}) in the massless limit.  However, as noted above,
gravitons can also mediate gluon-gluon contributions to lepton pair production
via $s$-channel exchange.  Such gluon initiated processes are a remarkable
consequence of this theory and have the potential to modify the Drell-Yan
spectrum in a unique manner.
Following an analogous procedure as outlined above for the four-fermion case, 
the matrix element for $gg\to\ell^+\ell^-$ via graviton exchanges is found to be
\bea
\label{ggll}
{\cal M}&=&{-4\lambda\over M_s^4}  
\bar f(p')[(p'-p)_\mu\gamma_\nu+(p'-p)_\nu\gamma_\mu]
f(p)\left\{k'_\alpha(k_\mu\eta_{\beta\nu}+k_\nu\eta_{\beta\mu})
+k_\beta(k'_\mu\eta_{\alpha\nu}+k'_\nu\eta_{\alpha\mu})\right.\nonumber\\
& & \left. -\eta_{\alpha\beta}(k'_\mu k_\nu+k_\mu k'_\nu)
+\eta_{\mu\nu}(k'\cdot k
\eta_{\alpha\beta}-k_\beta k'_\alpha)-k\cdot k'(\eta_{\mu\alpha}
\eta_{\nu\beta}+\eta_{\mu\beta}\eta_{\nu\alpha}\right\}\epsilon^\beta_g
(k')\epsilon^\alpha_g(k) \,,\nonumber\\
\eea
where the momentum flow is defined with both $k\,, k'$ flowing into the
vertex and $p\,, p'$ being outgoing and $\epsilon$ represents the gluon
polarization vector.  Because the graviton couplings and the
summation over the KK tower of states for $2\to 2$ processes are universal,
$\lambda$ is the same ${\cal O}(1)$ coefficient as in Eq. (\ref{ffff}).
This matrix element yields the $gg\to\ell^+\ell^-$ 
differential cross section for massless leptons
\be
\label{glue}
{d\sigma\over dz} = {\lambda^2\hat s^3\over 64\pi
M_s^8} (1-z^2)(1+z^2)\,,
\ee
which has a remarkably simple form.
The large parton luminosity
for gluons at higher energy colliders may also compensate for the $M_s^{-8}$
dependence.
Since this cross section is also even in $\cos\theta$,
the gluon-gluon contributions will only affect the total cross section and
not the forward-backward asymmetry.  Also note that the ambiguity in the
sign of $\lambda$ does not affect the gluon-gluon contributions
as they do not interfere with the $q\bar q$ initiated process.

\begin{figure}[t]
\centerline{
\psfig{figure=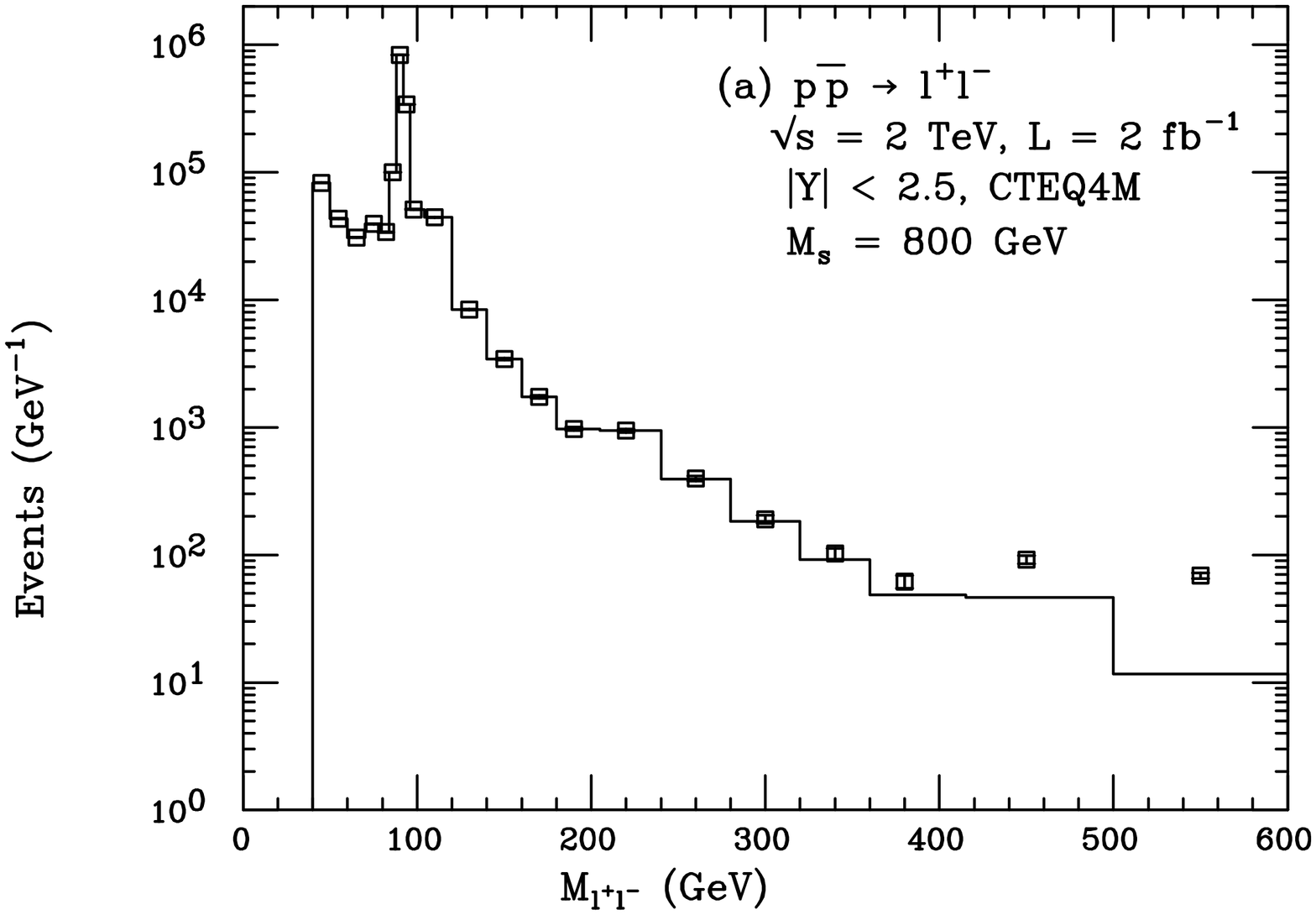,height=8.cm,width=8cm,angle=-90}
\hspace*{-5mm}
\psfig{figure=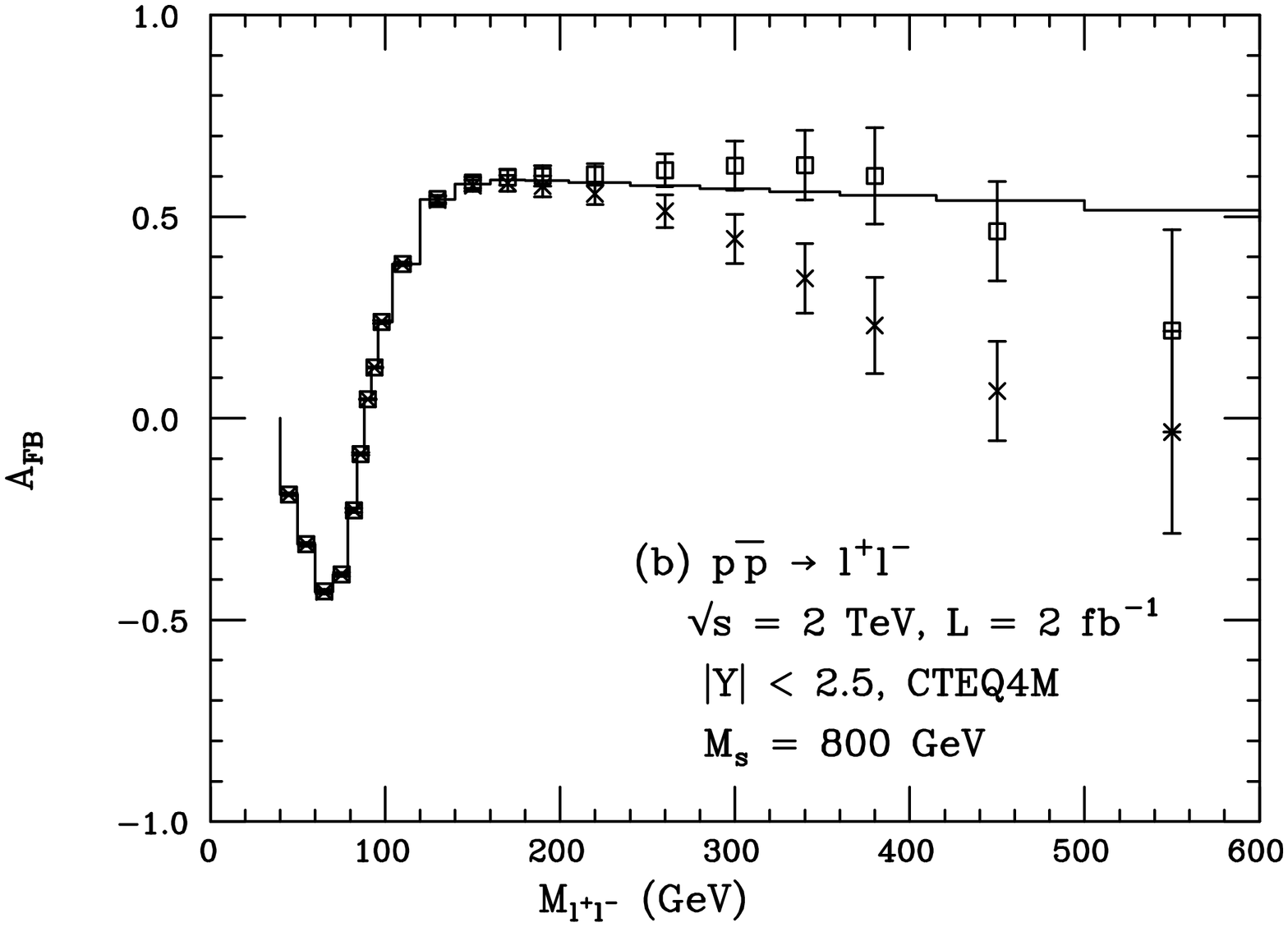,height=8.cm,width=8cm,angle=-90}}
\vspace*{-0.75cm}
\centerline{
\psfig{figure=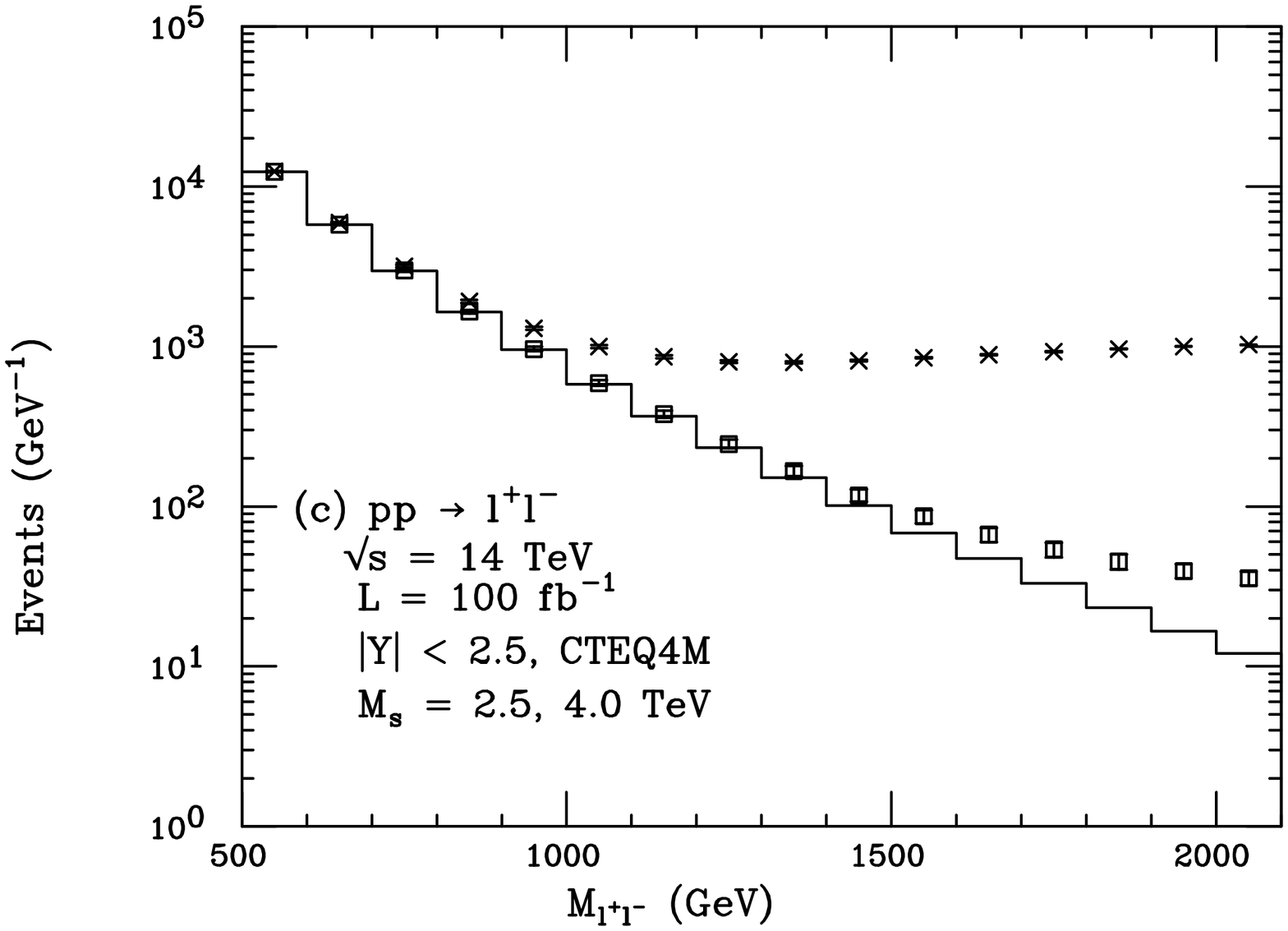,height=8.cm,width=8cm,angle=-90}
\hspace*{-5mm}
\psfig{figure=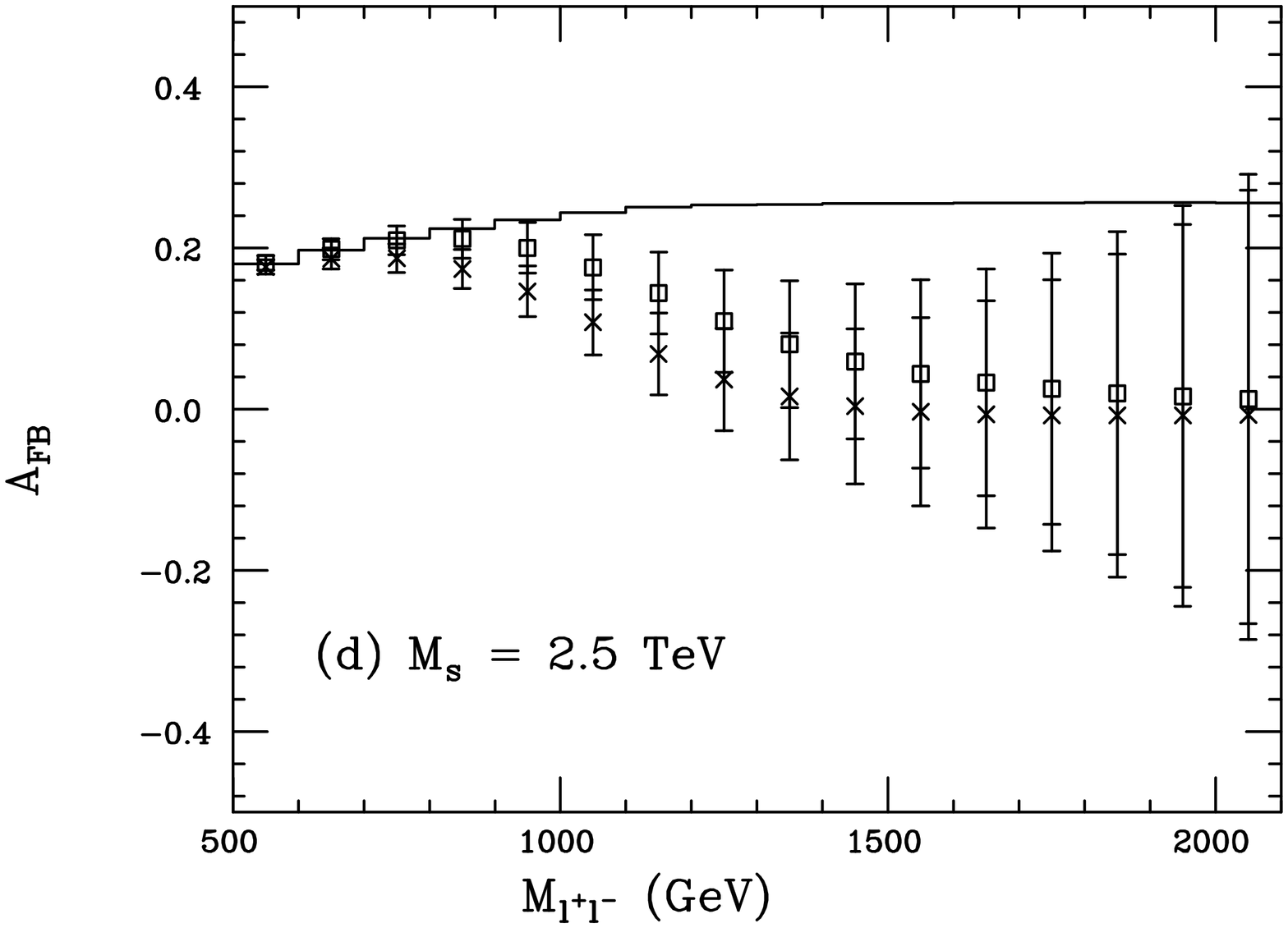,height=8.cm,width=8cm,angle=-90}}
\vspace*{-0.75cm}
\caption{Bin integrated lepton pair invariant mass distribution and
forward-backward asymmetry for Drell-Yan production at the Main Injector and
the LHC.  The SM is represented by the solid histogram.  The data points
represent graviton exchanges with (a) $M_s=800$ GeV and $\lambda=+1$ or $-1$,
(b) $M_s=800$ GeV and $\lambda=+1$ and $-1$, (c) $M_s=2.5$ and 4.0 TeV
and $\lambda=+1$ or $-1$, (d) $M_s=2.5$ TeV and $\lambda=+1$ and $-1$.}
\label{haddists}
\end{figure}

\begin{figure}[t]
\centerline{
\psfig{figure=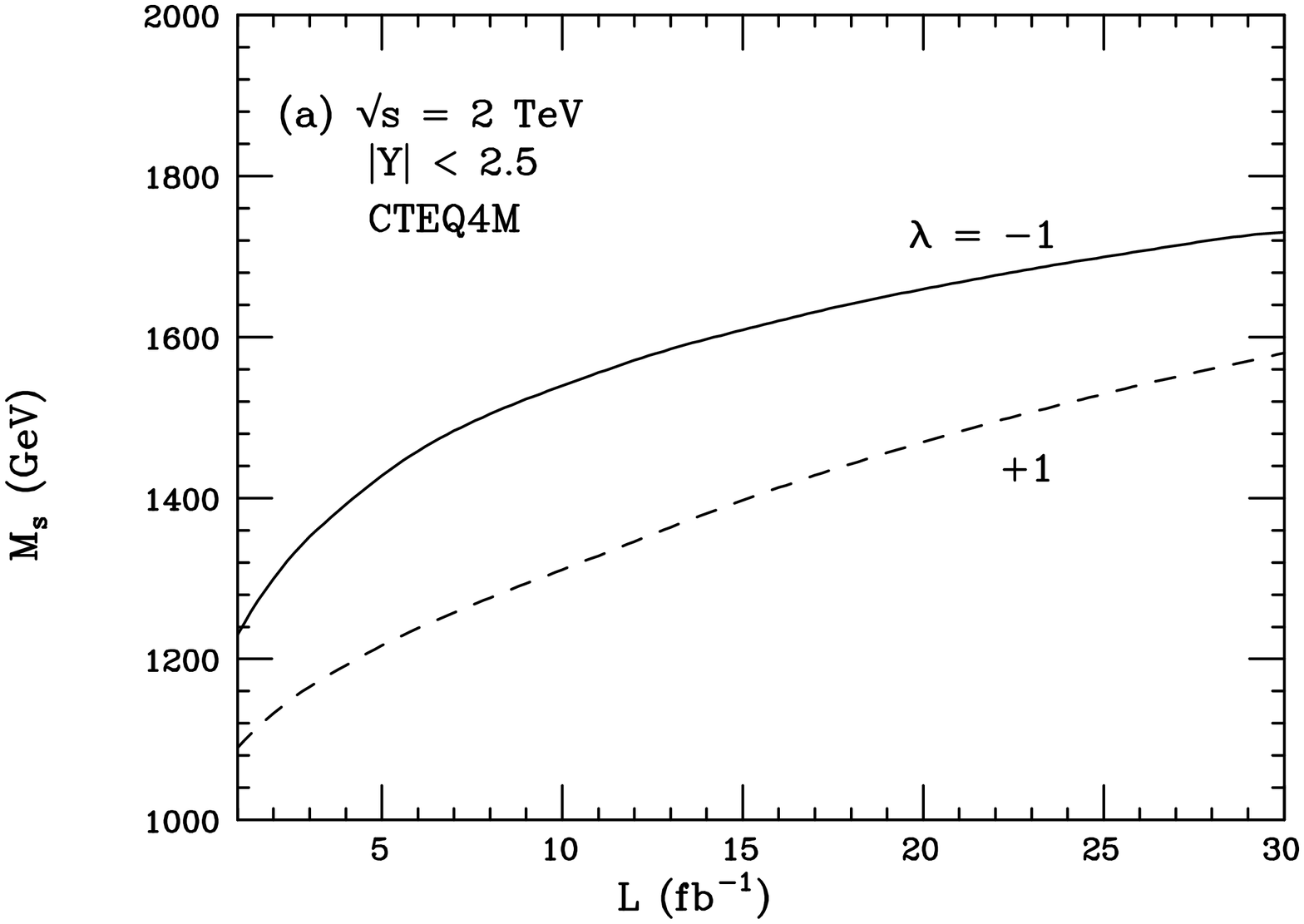,height=8.cm,width=8cm,angle=-90}
\hspace*{-5mm}
\psfig{figure=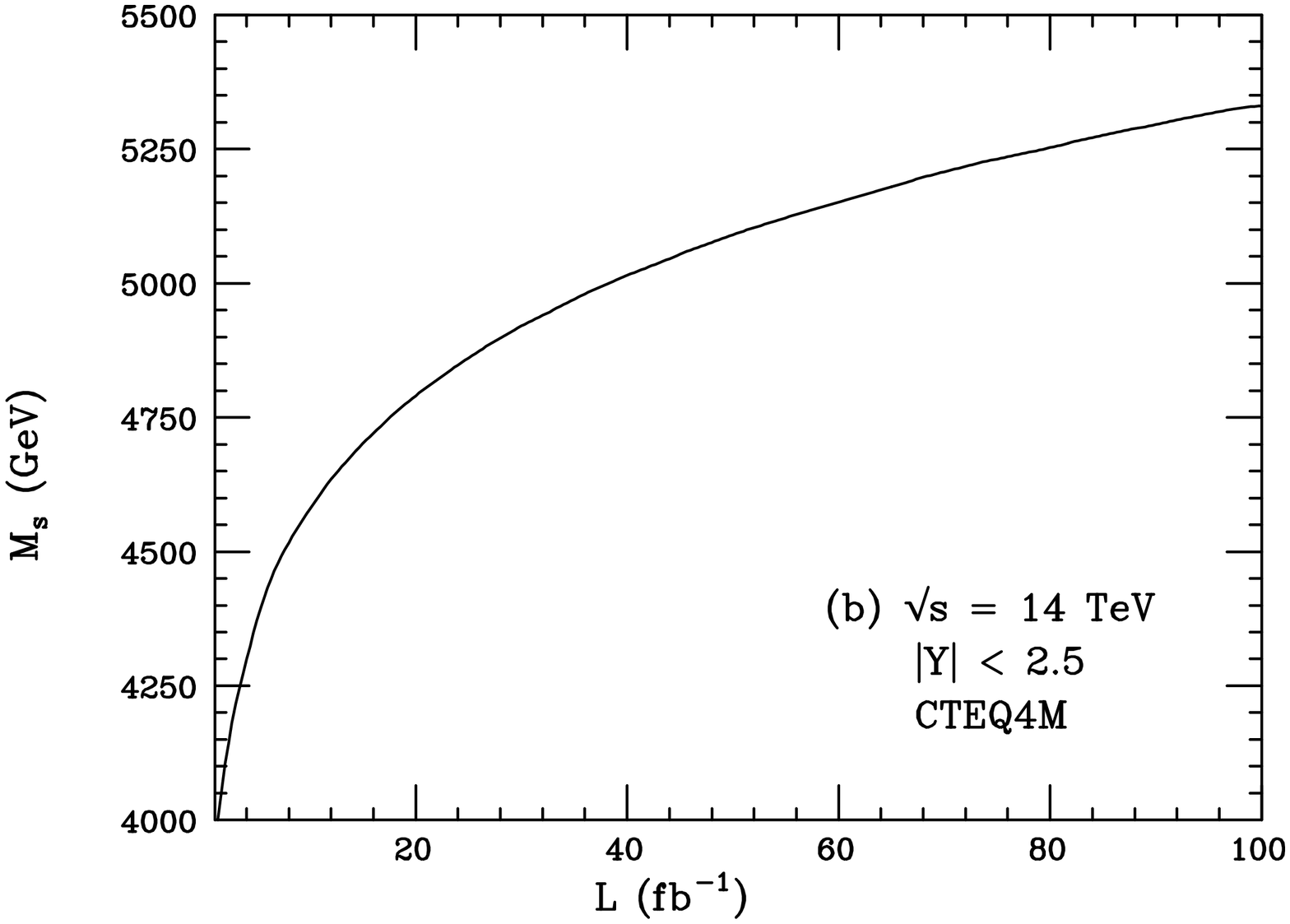,height=8.cm,width=8cm,angle=-90}}
\vspace*{-1cm}
\caption{95\% C.L. search reach for the string scale as a function of
integrated luminosity at the (a) Tevatron with the sign of $\lambda$ as
labeled and (b) LHC for either sign.}
\label{hadres}
\end{figure}

The bin integrated lepton pair invariant mass distribution and forward-backward
asymmetry $A_{FB}$ 
is presented in Fig. \ref{haddists} for the Tevatron Main Injector
and the LHC.  In each case the solid histogram represents the SM expectations,
and the `data' points include the graviton exchanges with the error bars
representing the statistics in each bin.  The rapidity cuts, parton density
parameterizations, and assumed integrated luminosity are as labeled, and we
have summed over electron and muon final states.  For
the Tevatron we show the sample case of $M_s=800$ GeV and the sign ambiguity
in $\lambda$ is visible in the forward-backward asymmetry.  For the LHC
we display the effects of a $M_s=2.5$ and 4 TeV string scale on the lepton
pair invariant mass spectrum (with the smaller string scale having the
larger effect), and again show $A_{FB}$ for both signs of the coefficient
taking $M_s=2.5$ TeV.  Since the graviton exchanges only affect 
the invariant mass distribution at order $\lambda^2/M_s^8$, we would expect
only minor modifications to this spectrum.  We see that this holds true for the 
Tevatron, however, large string scales do have a sizable effect
on the $M_{\ell\ell}$ spectrum at the LHC; this is due to the large gluon
luminosity at these center-of-mass energies.  The deviations in $A_{FB}$,
however, are not as pronounced at the LHC, whereas even the two cases 
$\lambda=\pm 1$ are statistically distinguishable from each other at the 
Tevatron for this sample case.  The resulting 95\% C.L. search reaches
are given in Fig. \ref{hadres} and and Table \ref{searchres} 
for both machines.  Here we see the effect of the
sign difference in the forward-backward asymmetry at the Tevatron, while the
LHC limits, which arise mainly from the $M_{\ell\ell}$ spectrum, are
independent of the sign.
We also find that present Tevatron data from Run I with 110\inpb\ of
integrated luminosity excludes a string scale up to 990 (930) GeV at 95\%
C.L. for $\lambda=-1(+1)$.

\vspace*{-0.5cm}
\nn

\vspace*{-0.5cm}
\nn

In conclusion, we have studied the indirect effects at high energy colliders
of a TeV string scale resulting from new large extra dimensions.  One
prediction of these theories is the existence of a Kaluza Klein tower of
massive gravitons, which can interact with the SM fields.  We derived the
form of these interactions and examined their
effect in the $2\to 2$ processes $\epem\to f\bar f\,, q\bar q\to \ell^+\ell^-$,
and $gg\to \ell^+\ell^-$ and found that present colliders can exclude a
string scale up to $\sim 1$ TeV and that future colliders can extend this
reach up to several TeV.  In addition, these constraints are essentially
independent of the number of extra dimensions as well as the details of the
full underlying theory.  Furthermore, we demonstrated that the angular 
distributions in \epem\ collisions uniquely reveal the spin-2 nature of the 
graviton exchanges and can be distinguished from other sources of new
physics for string scales close to the discovery limit and at a high confidence 
level.

These recent theories of low-scale quantum gravity are exciting, precisely
because they have numerous experimentally testable consequences.  The
phenomenology of these models is just beginning to be explored and we look
forward to the continued theoretical, phenomenological, and experimental
investigations of these theories.

\noindent{\bf Acknowledgements}
The author would like to thank Tao Han, Joe Lykken, and Tom Rizzo for 
discussions related to this work, 
and Nima Arkani-Hamed, Savas Dimopoulos, and
John Conway for their enthusiasm about this work.  After this work was
completed related material by Giudice \etal\cite{jimbo} appeared.


%
\def\IJMP #1 #2 #3 {Int. J. Mod. Phys. A {\bf#1},\ #2 (#3)}
\def\MPL #1 #2 #3 {Mod. Phys. Lett. A {\bf#1},\ #2 (#3)}
\def\NPB #1 #2 #3 {Nucl. Phys. {\bf#1},\ #2 (#3)}
\def\PLBold #1 #2 #3 {Phys. Lett. {\bf#1},\ #2 (#3)}
\def\PLB #1 #2 #3 {Phys. Lett. {\bf#1},\ #2 (#3)}
\def\PR #1 #2 #3 {Phys. Rep. {\bf#1},\ #2 (#3)}
\def\PRD #1 #2 #3 {Phys. Rev. {\bf#1},\ #2 (#3)}
\def\PRL #1 #2 #3 {Phys. Rev. Lett. {\bf#1},\ #2 (#3)}
\def\PTT #1 #2 #3 {Prog. Theor. Phys. {\bf#1},\ #2 (#3)}
\def\RMP #1 #2 #3 {Rev. Mod. Phys. {\bf#1},\ #2 (#3)}
\def\ZPC #1 #2 #3 {Z. Phys. C {\bf#1},\ #2 (#3)}


\begin{thebibliography}{99}

\bibitem{nima}
N. Arkani-Hamed, S. Dimopoulos, and G. Dvali, \PLB B429 263 1998 ;
hep-ph/9807344.
%
\bibitem{grexp}
S. Dimopoulos and G.F. Giudice, \PLB B379 105 1996 ; J.C. Long, H.W. Chan,
and J.C. Price, hep-ph/9805217.
%
\bibitem{tye}
I. Antoniadis, N. Arkani-Hamed, S. Dimopoulos, and G. Dvali, hep-ph/9804398;
G. Shiu and S.-H. Tye, hep-th/9805157; Z. Kakushadze and S.-H. Tye,
hep-th/9809147.
%
\bibitem{strings}
I. Antoniadis, \PLB B246 377 1990 ;
J. Lykken, \PRD D54 3693 1996 ; 
E. Witten, \NPB B471 135 1996 ;
P. Horava and E. Witten, \NPB B460 506 1996 , \ibid, {\bf B475}, 94 (1996);
E. Caceres, V.S. Kaplunovsky, I.M. Mandelberg, \NPB B493 73 1997 ;
K.R. Dienes, \PLB B197 139 1997 ;
K.R. Dienes, E. Dudas, and T. Gherghetta, hep-ph/9803466, hep-ph/9806292.
%
\bibitem{kkprod}
I. Antoniadis, K. Benakli, and M. Quiros; \PLB B331 313 1994 .
%
\bibitem{rs}
R. Sundrum, hep-ph/9805471.
%
\bibitem{gr}
M.J.G. Veltman, "Quantum Theory of Gravitation" in {\it Methods in Field
Theory}, Les Houches 1975, p. 265;
S. Weinberg, \PRD 138 B988 1965 ;
S.Y. Choi, J. Lee, S.S. Shin, and H.S. Song, \PRD D48 769 1993 ;
M.D. Scadron, {\it Advanced Quantum Field Theory and it's Applications Through
Feynman Diagrams}, (Springer-Verlag, 2nd ed, 1991);
P. van Nieuwenhuizen, \PR 68 121 1981 .
%
\bibitem{joe}
J. Lykken, talk presented at {\it Workshop on Physics and Detectors for
Future \epem\ Linear Colliders}, Keystone, CO, September 1998.
%
\bibitem{leptos}
J.L. Hewett and T.G. Rizzo, \PRD D56 5709 1997 .
%
\bibitem{tgr}
T.G. Rizzo, SLAC-PUB-7982, hep-ph/9811440.
%
\bibitem{djackson}
C.J.S. Damerall and D.J. Jackson, in Proceedings of the {\it 1996 DPF/DPB Summer
Study on New Directions for High Energy Physics}, Snowmass, CO, July 1996,
ed. D.G. Cassell \etal, (SLAC, Stanford, CA 1997) p. 442.
%
\bibitem{snow}
See, for example, T.G. Rizzo, in Proceedings of the {\it 1996 DPF/DPB Summer
Study on New Directions for High Energy Physics}, Snowmass, CO, July 1996,
ed. D.G. Cassell \etal, (SLAC, Stanford, CA 1997) p. 864 and p. 900.
%
\bibitem{jimbo}
G.F. Giudice, R. Rattazzi, and J.D. Wells, hep-ph/9811291;
S. Nussinov and R. Shrock, hep-ph/9811323;
E. Mirabelli, M. Peskin, and M. Perelstein, hep-ph/9811337;
T. Han, J. Lykken, and R.-J. Zhang, hep-ph/9811350.

\end{thebibliography}
\end{document}